\documentclass[10pt,aps,pra,twocolumn,superscriptaddress,floatfix,nofootinbib]{revtex4-2}

\usepackage{amsmath, amsfonts, amssymb}
\usepackage{bm,bbm, braket, accents}
\usepackage{graphicx}   

\usepackage[usenames,dvipsnames]{color}
\usepackage{lipsum} 
\usepackage{tikz}
\usetikzlibrary{quantikz}
\usepackage{svg}

\makeatletter
\def\@bibdataout@aps{%
 \immediate\write\@bibdataout{%
  @CONTROL{%
   apsrev41Control,author="08",editor="1",pages="0",title="0",year="1",eprint="1"%
  }%
 }%
 \if@filesw
  \immediate\write\@auxout{\string\citation{apsrev41Control}}%
 \fi
}%
\makeatother 

\newcommand{\dg}{^\dagger}
\newcommand{\smallfrac}[2]{\mbox{$\frac{#1}{#2}$}}
\newcommand{\half}{\smallfrac{1}{2}}

\newcommand\Tr{\mathrm{Tr}}
\newcommand{\ip}[2]{\langle{#1}|{#2}\rangle}
\newcommand{\op}[2]{\ket{#1}\!\bra{#2}}
\newcommand{\expt}[1]{\langle{#1}\rangle}

\newcommand{\triv}{\textsc{triv}}
\newcommand{\cat}{\textsc{cat}}
\newcommand{\bino}{\textsc{bin}}

\newcommand{\onerand}{\textsc{1-rand}}
\newcommand{\tworand}{\textsc{2-rand}}

\newcommand{\lbar}{\bar}

\newcommand{\Z}{\lbar Z}

\newcommand{\R}{\lbar R}

\usepackage[breaklinks=true]{hyperref}
\hypersetup{
  colorlinks   = true, 
  urlcolor     = blue, 
  linkcolor    = blue, 
  citecolor    = red 
}

\usepackage[normalem]{ulem}

\usepackage[capitalise]{cleveref} 
\crefformat{equation}{Eq.~(#2#1#3)} 
\crefformat{section}{Sec.~#2#1#3} 
\Crefformat{equation}{Equation~(#2#1#3)}
\crefformat{figure}{Fig.~#2#1#3}
\crefrangeformat{equation}{Eqs.~#3(#1)#4--#5(#2)#6}
\Crefformat{section}{Section~#2#1#3}

\usepackage{graphicx}

\begin{document}

\title{The performance of random bosonic rotation codes}

\author{Saurabh Totey}
\altaffiliation{These two authors contributed equally.}
\affiliation{Department of Physics, University of Colorado, Boulder, Colorado 80309, USA}
\affiliation{Department of Computer Science, University of Colorado, Boulder, Colorado 80309, USA}

\author{Akira Kyle}
\altaffiliation{These two authors contributed equally.}
\affiliation{Department of Physics, University of Colorado, Boulder, Colorado 80309, USA}

\author{Steven Liu}
\affiliation{Department of Aerospace Engineering, University of Colorado, Boulder, CO 80309}

\author{Pratik J. Barge}
\affiliation{Hearne Institute for Theoretical Physics, Department of Physics and Astronomy, Louisiana State University, Baton Rouge, Louisiana 70803, USA}

\author{Noah Lordi}
\affiliation{Department of Physics, University of Colorado, Boulder, Colorado 80309, USA}

\author{Joshua Combes }
\affiliation{Department of Electrical, Computer and Energy Engineering, University of Colorado, Boulder, Colorado 80309, USA}

\date{\today}


\begin{abstract}
  Bosonic error correcting codes utilize the infinite dimensional Hilbert space of a harmonic oscillator to encode a qubit.
  Bosonic rotation codes are characterized by a discrete rotation symmetry in their Wigner functions and include codes such as the cat and binomial codes.
  We define two different notions of random bosonic rotation codes and numerically explore their performance against loss and dephasing.
  We find that the best random rotation codes can outperform cat and binomial codes in a certain parameter regime where loss is large and dephasing errors are small.

\end{abstract}


\maketitle

\section{Introduction} \label{sec:intro}
Since \citet{Shannon1948}, error correction and randomness have been and continue to be intertwined~\cite{Gallager1973,BargForney2002,TruongCocco2023}. Similarly, in quantum information, random quantum codes~\cite{Holevo1998} have a long history of exploration, predominantly for qubits and qudits (e.g. \cite{Klesse2008,HaydenShorWinter2008,BrownFawzi2013}), and they are currently being actively explored \cite{FaistNezami2020,KongLiu2022}.

Bosonic codes encode qubits or qudits into a subspace of the quantum harmonic oscillator~\cite{GKP2001}. Bosonic error correcting codes have gained a lot of recent attention due to experimental demonstrations of logical fidelities exceeding physical fidelities~\cite{Sivak_GKP_2023,Ni2023}. Indeed, much of the theoretical progress has been spurred on by the rapid experimental progress in circuit QED~\cite{Ofek_2016,Hu_binomial_2019}, trapped ions~\cite{Fluhmann:2018aa,Fluhmann:2019aa,BurdSlichter2021,deNeeve2022}, and optics~\cite{Larsen_2021,Asavanant_2021,Sakaguchi_2023}. 
There are two major kinds of bosonic codes: those with a translation symmetry~\cite{GKP2001} (e.g. GKP codes), and those with a rotation symmetry~\cite{GrimCombBara20} (e.g. cat and binomial codes~\cite{Cochrane99,Ralph2003,Zaki,Michael16}). These codes have a structure induced by the translation or rotation symmetry and additionally they have a structure induced by the code family e.g. cat codes have a possion distribution in the number basis. Little has been done on bosonic random codes that are unsturctued in either way. However in 
 Ref.~\cite{ArzaniFerrini2019} the authors considered a continuous-variable bosonic random encoding for quantum secret sharing protocols,  while \cite{conrad_good_2023} introduced the randomized construction of GKP codes using the NTRU cryptosystem for a public key quantum communication scheme.

In this article, we construct two kinds of random bosonic codes that have the rotation symmetry strucutre but derive their fock amplitudes from haar random states. The codes have an $N$-fold rotation symmetry such that a phase space rotation of $\pi/N$ acts as a logical $Z$ gate. 
In \cref{sec:rot_codes}, we briefly describe rotation codes. Our random rotation codes are defined in \cref{sec:rand_rot_codes}. 
We introduce a two-parameter bosonic loss-dephasing channel in \cref{sec:err_chan}, which are typically the dominant errors for oscillator systems. In \cref{sec:code_perfom}, we compare the performance of our random codes to cat and binomial codes under these error channels using a numerically optimized recovery. We show that random codes can outperform both cat and binomial codes in certain regions of loss and dephasing. Building on previous work by  \citet{Leviant2022quantumcapacity}, we plot a 2D phase diagram of the best code as a function of loss and dephasing, which shows that, by and large, binomial codes are generally the most performant for the majority of realistic noise channels, but random codes can perform better under high loss and moderate dephasing.

\section{Rotation Codes} \label{sec:rot_codes}

Rotation-symmetric codes, or ``rotation codes'', are a type of bosonic code where discrete rotational symmetries in phase-space are utilized to protect against noise. Interestingly, the logical $Z$ gate for these codes is a rotation by $\pi/N$. The phase space rotation symmetry imposes some restrictions on the Fock coefficients of the code words. In order for a code to have a rotation symmetry degree of $N$, the $\ket{0_N}$ and $\ket{1_N}$ codewords must be representable as
\begin{subequations}\label{eq:logicalz_codewords}
\begin{align}
    \ket{0_N} ={}& \sum_{k=0}^\infty f_{2kN} \ket{2kN}, \label{eq:logicalz_0} \\
    \ket{1_N} ={}& \sum_{k=0}^\infty f_{(2k+1)N} \ket{(2k+1)N} \label{eq:logicalz_1} \, ,
\end{align}
\end{subequations}
where $\ket{k}$ is an eigenstate of the photon number operator $\hat n$. The amplitudes $f_{2kN}$ and $f_{(2k+1)N}$ are the only things that specify a class of rotation codes, so a class of codes is determined by the functional form of these coefficients. The class of rotation code does not restrict the possible rotation symmetry degrees.

The dual-basis codewords, $\ket{\pm_N}$, are constructed as usual via superpositions of the computational basis codewords. $\ket{\pm_N} = \frac{1}{\sqrt{2}}(\ket{0_N} \pm \ket{1_N})$, yielding Fock space representations
\begin{subequations} \label{eq:logicalx_codewords}
\begin{align} 
    \ket{+_N} = & \frac{1}{\sqrt{2}}\sum_{k=0}^\infty  f_{k N} \ket{k N}\label{eq:logicalxplus},\\
    \ket{-_N} = & \frac{1}{\sqrt{2}} \sum_{k=0}^\infty  (-1)^k f_{k N} \ket{k N}\label{eq:logicalxminus}.
\end{align}
\end{subequations}
Both $\ket{\pm_N}$ have support on the full set of $\ket{k N}$ Fock states. It is these dual basis codewords that have an $N$-fold rotational symmetry, while the $Z$ basis code words in \cref{eq:logicalz_codewords} have a $2N$-fold rotation symmetry. 

 A general logical state $\ket{\psi}_L = \alpha \ket{0_N} + \beta \ket{1_N}$ is the $+1$ eigenstate of the stabilizer
\begin{equation}\label{eq:Z_N}
     \R_N\equiv \exp\! \left[ i \frac{2\pi}{N} \hat n  \right] \, .
\end{equation}
It turns out that the logical $\Bar{Z}$ gate is given by
\begin{equation}
 \Z_N\equiv \sqrt{\hat R_N} = \hat R_{2N}   
\end{equation}
which is a rotation by $\pi/N$ such that $ \Z_N\ket{\pm_N} = (\pm 1)\ket{\pm_N}$. The corresponding logical $\Bar{X}$ gate is code dependent. However, rotations about $\Bar{Z}$ by fractional angles can be constructed in a code independent way~\cite{GrimCombBara20,MarinoffBushCombes2023}.

There are two natural errors for a rotation code to correct. The first kind are shift errors, where a number state $\ket{n}$ gets shifted by $g$ i.e. $\ket{n}\mapsto \ket{n \pm g}$ where $g$ is some integer. The second kind are rotation errors which are induced by the operator $e^{i\phi \hat n}$. The set of shift and rotation errors that are mutually correctable for an order-$N$ ideal phase code are given by~\cite{GrimCombBara20,Yingkai2021,MarinoffBushCombes2023}
\begin{subequations} \label{eq:code_distancey}
\begin{align}
    g &\in [0, N) \quad {\rm (number\ errors)} \\
    \phi &\in [0, \pi/N) \quad {\rm (phase\ errors)}.
\end{align}
\end{subequations}
There is a trade-off between the ability to correct number shift and phase errors. For example, a higher rotation symmetry degree, i.e. $N$, allows for rotation codes to protect against shift errors more effectively, but that, in turn, adversely affects the ability to protect against phase errors. The consequences of this trade-off are made apparent in \cref{sec:code_perfom}.

\section{Random Rotation codes} \label{sec:rand_rot_codes}

We start by considering rotation codes where the amplitudes of the code words are generated with some random processes. We describe two kinds of random rotation codes. Both of the following random generation methods rely on using Haar random unitary matrices~\cite{Spengler2012} applied to arbitrary states to get random states. A $(K+1)\times (K+1)$ Haar random unitary is denoted as
\begin{equation}\label{eq:Haar_U}
    \mathbb{U}_{K+1} \in {\rm SU}(K+1) \, .
\end{equation}
The restriction to $K+1$ invokes a photon number cutoff of $K$, so the unitary acts on Fock states from vacuum $\ket{0}$ through $\ket{K}$. It should be noted that there are simple algorithms to generate Haar random samples from the special unitary group~\cite{Mezzadri2007}. All random states in this work were generated by applying the Haar random unitary to the ground state of the quantum harmonic oscillator i.e. $\ket{0}$.

\subsection{Random codes with one random primitive state}\label{sec:one_rand_codes}

\begin{figure}[h]
\centering
\includegraphics[page = 2,width=0.99\columnwidth]{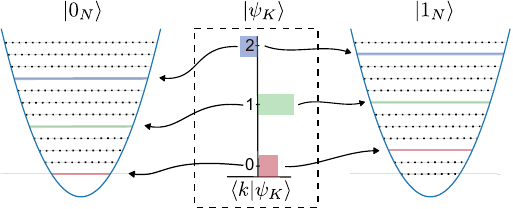}
\caption{Using one randomly generated primitive for both logical zero and logical one codewords. In the center, we plot the real part of the fock amplitudes of the randomly generated state i.e. $\Re(\ip{k}{\psi_K})$. In practice, we allow these weights to be complex. This state gets expanded separately into the $\ket{0_N}$ and $\ket{1_N}$ spaces. Importantly, the primitive state only has as many dimensions as will be used in the codewords.
}\label{fig:one_rand_prim_codes}
\end{figure}

Here we construct codes where the logical codewords $\ket{0_N}$ and $\ket{1_N}$ use the same random state but are offset in Fock space. For example, if the codewords are allowed to use 7 levels of the harmonic oscillator, then a 7-dimensional random state is generated. The random state is then `expanded' into the code words by having the used levels of the harmonic oscillator match the corresponding values of the random state vector. This process is illustrated in \cref{fig:one_rand_prim_codes}.

In general, we can consider a $K+1$-dimensional random state that we call a ``primitive''
\begin{align}\label{eq:1_rand_prim}
    \ket{\psi_K}= \mathbb{U}_{K+1}\ket{0}= \sum_{k=0}^K \psi_k \ket{k} \, ,
\end{align}
where $\ket{0}$ is the vacuum and $\ket{k}$ is the k'th Fock state. This state gets ``expanded'' into the code words
\begin{subequations}
\begin{align}
    \ket{0_N} ={}& \sum_{k=0}^K \psi_{k} \ket{2Nk}\, ,  \\
    \ket{1_N} ={}& \sum_{k=0}^K \psi_{k} \ket{(2k+1)N}  \, ,
\end{align}
\end{subequations}
where $\psi_k = \ip{k}{\psi_K}$ where $\ket{\psi_K}$ is defined in \cref{eq:1_rand_prim}. Even though the amplitude is the same on both code words, the states are orthogonal (i.e. $\ip{0_N}{1_N}=0$) due to being constructed on non-overlapping Fock states. Moreover, we can approximately control the average photon number in the code space by varying the cutoff, $K$. If we naively assume flat amplitudes on average for the random primitive state, the expected value of photon number for the code space is $\expt{\hat n} = N(2K+1)/2$.

\subsection{Random Codes with two random primitive states}\label{sec:TwoExpand}

\begin{figure}[th]
\centering
\includegraphics[page = 3,width=0.99\columnwidth]{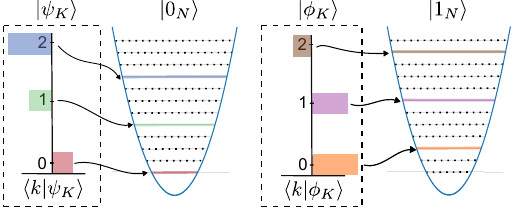}
\caption{Using two randomly generated primitive states for logical zero and logical one codewords separately. Each logical codeword draws from its own primitive state, and each primitive state is exactly as large as is needed by the codeword. Again, we allow the weights on each Fock state to be complex, but plot only the real part for simplicity. 
}\label{fig:two_rand_prim_codes}
\end{figure}

These codes use two random states to construct the code, one for $\ket{0_N}$ and one for $\ket{1_N}$. For this reason we consider two $K+1$-dimensional random primitives
\begin{align}
\ket{\psi_K}= \mathbb{U}_{K+1}\ket{0}= \sum_{k=0}^K \psi_k \ket{k} \, \\
\ket{\phi_K}= \mathbb{V}_{K+1}\ket{0}= \sum_{k=0}^K \phi_k \ket{k} \, ,
\end{align}
where $\mathbb{U}_{K+1}$ and $\mathbb{V}_{K+1}$ are $(K+1)\times (K+1)$ random unitaries. In principle, we could consider different cutoffs for the different states, but for simplicity we do not. These states get expanded into the codewords
\begin{subequations}
\begin{align}
    \ket{0_N} ={}& \sum_{k=0}^K \psi_{k} \ket{2Nk},  \\
    \ket{1_N} ={}& \sum_{k=0}^K \phi_{k} \ket{(2k+1)N}  \, , 
\end{align}
\end{subequations}
where $\psi_k$ and $\phi_k$ are amplitudes generated by {\em different} random unitaries. The process of constructing these codes is depicted in \cref{fig:two_rand_prim_codes}. Note that the codes are orthogonal due to the non-overlapping support on the Fock states despite $|\psi_K\rangle$ and $|\phi_K\rangle$ generally not being orthogonal.

\begin{figure}[th]
    \centering
    \includegraphics[page=12,width=0.99\columnwidth] {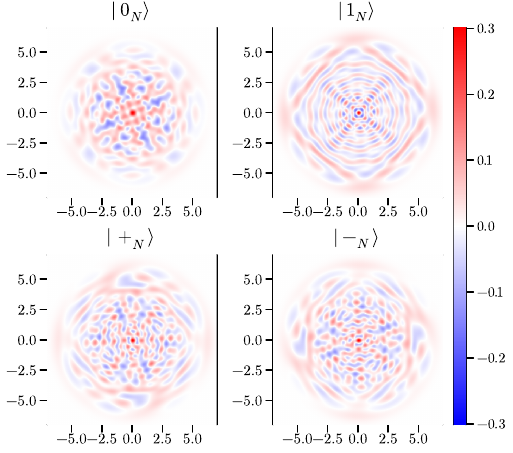}
    \caption{Wigner functions of a random code with two primitives described in \cref{sec:TwoExpand} with rotation symmetry of $N=2$ and a cut-off of  $K=5$. The $\ket{\pm_N}$ states exhibit a 2-fold symmetry, while the $Z$ basis codewords have a $2N$ symmetry i.e. a $4$-fold rotation symmetry.}
    \label{fig:eg_rand_code}
\end{figure}

The Wigner functions of an example random code with two random primitive states is depicted in \cref{fig:eg_rand_code}. Notice that the Wigner function for $\ket{0_N}$ and $\ket{1_N}$ are different, which is normal in these kinds of codes. This is unlike the codes presented in \cref{sec:one_rand_codes}, which will have similar looking Wigner functions for $\ket{0_N}$ and $\ket{1_N}$.

It is worthwhile to pause and think about the set of codes from which we are sampling. For a fixed value of rotation symmetry $N$ and Fock cutoff $K$, we will call the set of all possible codes $\mathbb{S}(N,K)$.  Random codes with one primitive sample a subset of the space of all possible codes, but we conjecture that the random codes with two primitives sample the complete space of $\mathbb{S}(N,K)$. This means that any possible code that shares this symmetry and cutoff could be generated by random codes with two primitives, even binomial and cat codes, although this would be incredibly unlikely for any single trial. We make an attempt to sample $\mathbb{S}(N,K)$ uniformly by generating states from the Haar measure.

\section{Error Channels}\label{sec:err_chan}

The dominant physical error channels in most bosonic systems are loss and dephasing.
The noise channel we consider consists of simultaneous loss and dephasing Specifically, the noise channel, $\mathcal N(\hat\rho)$, is the solution to the master equation
\begin{equation}\label{eq:loss_dephasing_me}
    \dot {\hat{\rho}} = \kappa_l \mathcal D[\hat a] \hat\rho + \kappa_\phi \mathcal D[\hat n] \hat\rho\;,
\end{equation}
where $\mathcal D[\hat L] \hat\rho = \hat L \hat\rho \hat L\dg - \half \hat L\dg \hat L \hat\rho - \half \hat\rho \hat L\dg \hat L$.
Rather than numerically integrating Eq.~\ref{eq:loss_dephasing_me} or finding a Kraus-operator representation, we exponentiate the superoperator representation of the Lindbladian. That is, we exponentiate $\mathcal{L}[\hat{L}]$, which is the superoperator representation of $\mathcal{D}[\hat{L}]$, which is given as $\mathcal{L}[\hat{L}] = \hat L^* \otimes \hat L - \half \hat I\otimes \hat L\dg \hat L - \half (\hat L \dg \hat L)^T \otimes \hat I$. We can then take advantage of the algorithm for sparse matrix exponentiation in Refs.~\cite{Hogben11,Kuprov11} to efficiently solve Eq.~\ref{eq:loss_dephasing_me} for some evolution time $t$, thereby giving the noise channel as
\begin{equation}\label{eq:loss_dephasing_ch}
\mathcal{N}(\hat{\rho}) = e^{t(\kappa_l \mathcal L[\hat{a}] + \kappa_\phi \mathcal L[\hat{n}])} \hat{\rho}\;.
\end{equation}
Thus $\kappa_l t$ and $\kappa_\phi t$ represent the unitless strengths of the simultaneous loss and dephasing rates.

\section{Code performance}\label{sec:code_perfom}

\begin{figure*}[!ht]
    \centering
    \includegraphics[page = 6,width=0.99\textwidth]{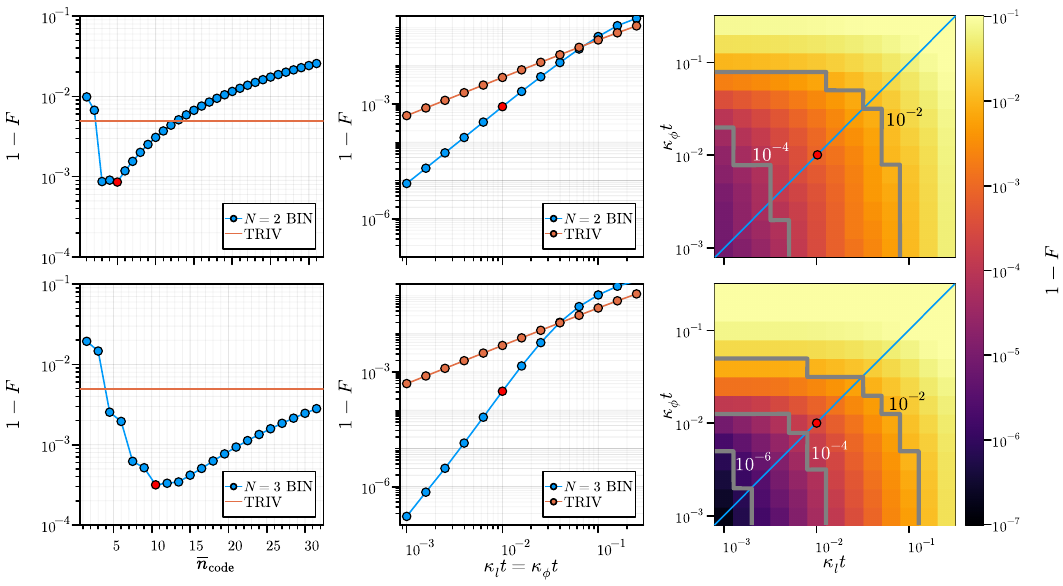}
    \caption{
    Performance of $N=2$ ({Row 1}) and $N=3$ ({Row 2}) binomial codes.
    {\bf Column 1.} The performance (logical channel infidelity) of binomial codes as a function of the average photon number of the code space projector when  $ \kappa_l t = \kappa_\phi t = 10^{-2}$. The minimum of this curve is plotted as a single red marker. 
    {\bf Column 2.} Code performance as a function of equal loss  and dephasing $\kappa_l t = \kappa_\phi t$. Notice the minimum from column 1 shows up as a single point for $ \kappa_l t = \kappa_\phi t = 10^{-2}$.
    {\bf Column 3.} Code performance as a function of dephasing and loss. Note that the diagonal part of this plot is the data from column 2 where $\kappa_l t = \kappa_\phi t$.   
    }    \label{fig:bin_N2_N3_code_compare}
\end{figure*}

In this section, we explore the performance of single-mode bosonic rotation codes subject to loss and dephasing noise using the methods described in \cref{sec:err_chan}.
In \cref{sec:perform_cat_n_binomial} we compare the performance of known rotation codes, namely cat and binomial codes~\cite{Cochrane99,Ralph2003,Michael16,Zaki}. 
In \cref{sec:perform_rand_codes} we explore the performance of our new random rotation codes. 
In \cref{sec:code_comparison} we compare the performance of cat and binomial codes to random rotation codes. We now briefly describe how these comparisons are performed.

Rather than assessing the performance of a particular error correction scheme, we use a semidefinite program (SDP) to compute the recovery, $\mathcal R^\text{SDP}$, that optimizes channel fidelity~\cite{FlectcherShorWin2007}. The channel fidelity of a channel $\mathcal E$ to the identity channel is 
\begin{equation}\label{eq:chan_fid}
    F(\mathcal I, \mathcal E) = \frac{1}{d(d+1)}\left (  \Tr\left[\sum_k \hat M_k^\dagger \hat M_k\right ] + \sum_k |\Tr[\hat M_k]|^2\right )\, ,
\end{equation}
where $\{ \hat M_k\}$ are the Kraus operators for $\mathcal{E}$ and $d$ is the dimension of the Hilbert space~\cite{Pedersen_fidelity_2007}.

Thus the total quantum channel we consider is
\begin{equation}\label{eq:ec:totalchannel}
    \mathcal D_{\rm dec} \circ \mathcal R^\text{SDP} \circ \mathcal N \circ \mathcal S,
\end{equation}
where
$\mathcal{S}: \mathcal{H}_2 \to \mathcal{H}_\mathbb{N}$ is the ideal encoding map for a given code,
$\mathcal N: \mathcal{H}_\mathbb{N} \to \mathcal{H}_\mathbb{N}$ is the noise map (\cref{eq:loss_dephasing_ch}), $\mathcal R^\text{SDP}: \mathcal{H}_\mathbb{N} \to \mathcal{H}_\mathbb{N}$ is the optimal recovery that maximizes the channel fidelity~\cite{FlectcherShorWin2007}, and finally $\mathcal D: \mathcal{H}_\mathbb{N} \to \mathcal{H}_2$ is the ideal decoding map for a given code.
The encoder superoperator is given by $\mathcal S[\hat S]\hat\rho = \hat S \hat\rho \hat S\dg$ where $\hat S = \ket{0_N}\bra 0 + \ket{1_N}\bra 1$ while the decoder is given by $\mathcal D_{\rm dec}[\hat S^\dagger]\hat\rho$.
Thus the total channel is a logical channel with qubit input and qubit output.

To this end, we compute the channel fidelity (to the identity channel) for the channels in \cref{sec:err_chan} for different rotation codes $\mathcal C \in [\cat, \bino, \onerand, \tworand]$ and different rotation symmetries $N\in [2,3,4]$. 
As we are considering a logical channel with qubit input and qubit output, $d=2$ in \cref{eq:chan_fid}. Moreover, we prefer the infidelity (i.e. $1-F$), as it is a measure of error.

When comparing the performance of codes, it has been customary~\cite{albert2018} to rank codes by the average photon number of the code space. This is done by defining the code space projector as
\begin{equation}
    \hat P_{\rm code} = \op{0_N}{0_N} + \op{1_N}{1_N}  \, ,
\end{equation}
so then the average photon number in the code space is
\begin{equation}\label{eq:avg_num_code}
    \bar{n}_{\rm code} =\frac{1}{2} \Tr[\hat{n} \hat P_{\rm code}] \, .
\end{equation}
The other important benchmark is to see when encoding into a code results in a lower channel infidelity than the ``naive'' encoding. The naive encoding is called the trivial code, and it is where the logical zero and one codewords are respectively the ground and first excited state of the harmonic oscillator,
\begin{equation}
    \ket{0_\triv} = \ket{0}, \quad {\rm and} \quad \ket{1_\triv} = \ket{1}\,.
\end{equation}
This is a useful comparison when the loss channel is involved, as states with higher photon number e.g. $\ket{K}$ will decay at a rate of $\kappa_l K$ which is much faster than that of $\ket{1}$ which decays at a rate $\kappa_l$. That is, we expect the \triv\ encoding to be good.

When we present diagrams of code performance as a function of $\kappa_l t$ and $\kappa_\phi t$,  the plots will be log-space with 5 points per decade and the range is 
\begin{equation}
 \kappa_l t, \kappa_\phi t \in[10^{-3}, 0.25]   
\end{equation}
We cut off infidelities below $1\times 10^{-7}$  due to the trade off between tolerances and runtime of the SDP solver.

\subsection{Exploration of cat and binomial}\label{sec:perform_cat_n_binomial}

There are several rotation codes that we could study: polygon codes, squeezed cat codes, Pegg-Barnett codes etc. We examine the two most popular rotation codes, cat codes~\cite{Cochrane99,Ralph2003,Zaki} and binomial codes~\cite{Michael16}, because they have the most theoretical and experimental work about them.

The Fock-grid coefficients  $f_{kN}$ from \cref{eq:logicalz_codewords} for these codes are
\begin{subequations}
\begin{align}
f_{kN}&=\sqrt{ \frac{1}{2^{K-1}} \binom{K}{k} } & {\rm (\bino)}\label{eq:bino}\\
f_{kN}&=\sqrt{ \frac{2}{\mathcal{N}_i} } \frac{e^{-|\alpha|^2/2} \alpha^{kN}}{\sqrt{(kN)!}} & {\rm (\cat)}
\end{align}
\end{subequations}
where, for cat codes, $\mathcal{N}_i$ is the Fock-space normalization factor such that $\mathcal{N}_0$ is used for even values of $k$ and $\mathcal{N}_1$ is used for odd values of $k$.

In row 1 of \cref{fig:bin_N2_N3_code_compare}, we plot the channel fidelity of the $N=2$ binomal code while in row 2, we plot the $N=3$ codes. In column 1 we sweep the code parameter $K$ in \cref{eq:bino} which controls the average photon number of the code space. A generic feature is that there is typically an optimal value of $K$ for a given noise value. Also, note that only some values of $K$ result in channel infidelities below the trial encoding. The filled red point indicates the point with the smallest channel infidelity. 

In column 2 of \cref{fig:bin_N2_N3_code_compare}, we plot the best-performing code (optimized over the binomial code parameter $K$) as a function of noise strength for the case where loss is equal to dephasing i.e. $\kappa_l t = \kappa_\phi t$. Again we are interested in when the codes are beating the \triv\ infidelity, which appears to be $5\times10^{-2}$ for $N=2$ and $7\times10^{-2}$ for $N=3$ codes. The optimal point from the previous plot now appears as one point on these plots.

In column 3 of \cref{fig:bin_N2_N3_code_compare}, we plot the best-performing code as a function of loss $\kappa_l t$ and dephasing $\kappa_\phi t$. The optimal codes from column 2 are now the diagonal line along this plot. A general trend we note is that, as either $\kappa_\phi t$ or $\kappa_l t$ decrease, the infidelity decreases. That is to say, as the noise strength decreases, the code performance increases.

Cat codes have similar performance as these binomial codes and the equivalent plot for cat codes can be found in the \cref{apx:more_plots} see \cref{fig:cat_N2_N3_code_compare}. In \cref{sec:code_comparison}, we directly compare \bino\ and \cat\ codes.


\begin{figure}[h]
\centering
\includegraphics[page = 4,width=0.95\columnwidth]{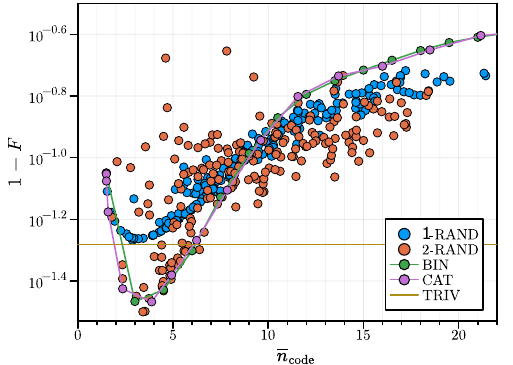}
\caption{Channel infidelity as a function for photon number for  $ \kappa_l t = 10^{-0.8}$ and $\kappa_\phi t = 10^{-1.8}$ and $N=3$. Binomial, and cat codes are shown as solid lines. While both \tworand\ codes and \onerand\ codes outperform binomial and cat codes at high photon number, the best performance from any code is achieved by a \tworand\ code for this noise channel. 
}\label{fig:rand_code_do_good}
\end{figure}

\begin{figure}[h]
\centering
\includegraphics[page = 5,width=0.95\columnwidth]{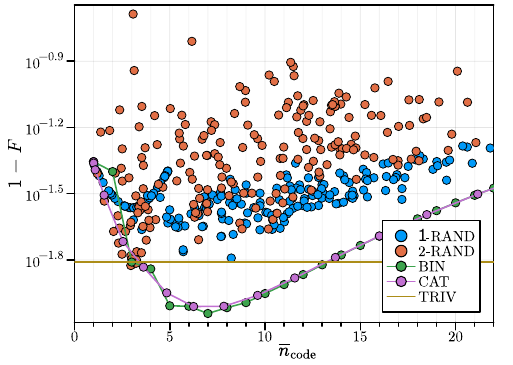}
\caption{
Channel infidelity as a function for photon number for  $ \kappa_l t = 10^{-1.8}$ and $\kappa_\phi t = 10^{-1.2}$ and $N=2$. Binomial, and cat codes are shown as solid lines. For this noise channel the random codes only rarely beat the trivial encoding and never beat the most performant binomial or cat codes. 
}\label{fig:rand_code_do_bad}
\end{figure}

\subsection{Exploration of random codes}
\label{sec:perform_rand_codes}

In this section, we begin to explore the performance of the random codes described in \cref{sec:rand_rot_codes}.
\cref{fig:rand_code_do_good} and \cref{fig:rand_code_do_bad} show the performance of these random codes for two different noise channels, one with large loss and moderate dephasing (\cref{fig:rand_code_do_good}), and one with large dephasing and moderate loss (\cref{fig:rand_code_do_bad}). These plots provide some basic intuition for the performance of random codes.

The first interesting feature from  \cref{fig:rand_code_do_good} is that, for certain noise channels where loss dominates dephasing, we see \tworand\ codes that outperform binomial codes, even with a modest number of random trials. For certain average photon numbers, almost all random codes perform better than both binomial and cat codes. This is explored more in \cref{sec:code_comparison} and in particular \cref{fig:compare_rand_bin}.

The second interesting feature in \cref{fig:rand_code_do_good} is that, for channels with low dephasing, the infidelity of \onerand\ codes appear to follow a smooth curve relating average photon number to performance. Further numerical studies (not presented here) suggest that, in the presence of no dephasing, the performance of \onerand\ codes depends solely on average photon number despite the random coefficients of the code.

Channels with low loss and high dephasing, such as the one used in \cref{fig:rand_code_do_bad}, have poor performance from both types of random codes. This is evident because only very few random codes beat the trivial code, with the exception of a few at $\bar{n}_{code} \approx 3$. Interestingly, the \onerand\ codes seem to be performing statistically better than the \tworand\ codes.
For many channels, especially those where heuristics imply that structured codes like cat and binomial codes are very good, generating a random code that outperforms these existing codes should be very rare.

The number of trials used in these simulations is insufficient to make claims about average performance or the limits of performance. Naively, one might expect that the best possible \tworand\ codes must perform as good or better than both binomial and cat codes for any channel because the set of codes that are randomly sampled include the known codes. We caution that randomly recreating known codes is very unlikely. This suggests it may also be possible to increase the likelihood of generating these exceptional random codes by sampling a distribution other than the Haar measure. This is left as future work.


\subsection{Code comparison}\label{sec:code_comparison}

In this section, we compare rotation codes in the set $\mathcal C \in [\cat, \bino, \onerand, \tworand]$ with respect to their ability to correct errors and their optimal code parameters, namely average photon number and rotational symmetry.

\begin{figure*}[ht]
    \centering
    \includegraphics[page = 8, width=0.99\textwidth]{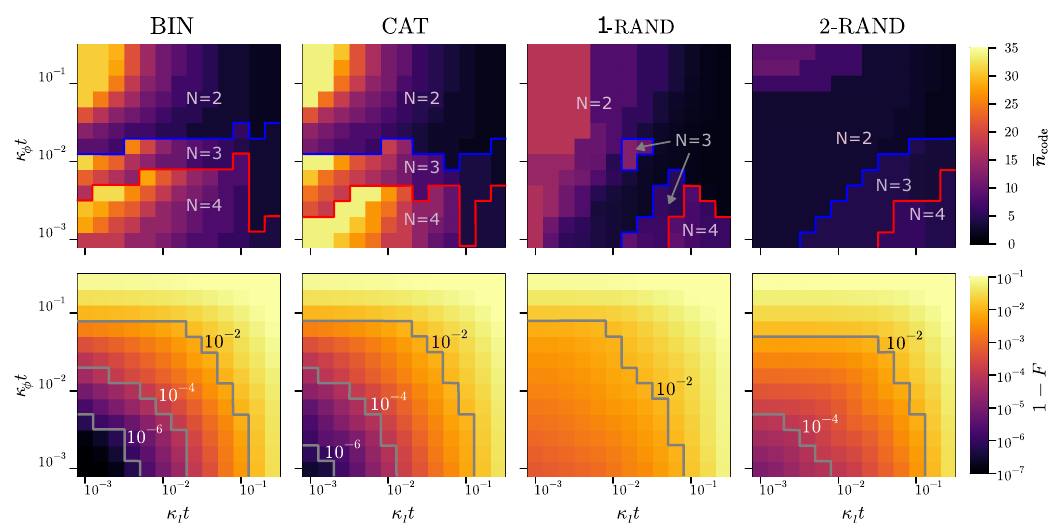}
    \caption{For each class of codes (\bino, \cat, \onerand, \tworand) and each noise channel (parameterized by $\kappa_\phi t$ and $\kappa_l t$), we find the optimal code parameters and plot the performance of the code as the channel infidelity. {\bf Row 1} The best performing average photon number of the code space projector (colorbar) and rotation symmetry (region) are plotted as a function of noise channel parameters. {\bf Row 2} Channel infidelity, $1-F$, of the best performing code as a function of loss and dephasing. 
    }    \label{fig:code_compare_and_sweep}
\end{figure*}

\cref{fig:code_compare_and_sweep} provides a side-by-side comparison of all four codes across many values of both loss and dephasing. In row 1 we plot the average photon number of the code space projector, i.e. \cref{eq:avg_num_code}, for the best code as a function of dephasing $\kappa_\phi t$ and loss $\kappa_l t$. One trend that stands out is that high loss encourages low photon number codes. At about $\kappa_l t \approx 4\times 10^{-2}$ we see the photon numbers starting to increase. The best ``structured'' codes, i.e. \bino\ and \cat\ codes, have increasing photon as the noise decreases. While the \onerand\ and \tworand\ codes seem to have generically have lower photon numbers.

The overlay on row 1 indicates the rotation symmetry of the best code for $N\in[2,3,4]$. 
This overlay reveals another apparent trend for the structured codes i.e. \bino\ and \cat\ codes. 
Notice the almost horizontal striation at  $\kappa_\phi t \approx 2\times 10^{-2}$ that appears because the optimal code switches from an $N=2$ to an $N=3$ code which necessitates a higher photon number. A similar diagonal feature can be seen at $\kappa_\phi t \approx 2.5\times 10^{-3}$, which occurs when the optimal code becomes $N=4$. The \onerand\ and \tworand\ codes seem to have more diagonal striations, this could be due to insufficient sampling of higher photon number random codes. Naively this may imply that the set of all performant codes does not scale with photon number as quickly as the set of all possible codes.

In row 2 of \cref{fig:code_compare_and_sweep}, we plot the channel infidelity of the best code. The channel infidelity is presented as a function of dephasing $\kappa_\phi t$ and loss $\kappa_l t$. We see, as expected, that as the noise decreases, the error (infidelity) decreases. Generally, binomial codes outperform all the other codes, but cat codes have a somewhat comparable behavior. This is because binomial codes were designed~\cite{Michael16} to correct a finite number of loss and dephasing events. Thus we expect them to be good codes for loss and dephasing channel. From the infidelity contours, it is evident that the \onerand\ and \tworand\ codes don't perform well in the low noise regime when compared to the structured codes.

\begin{figure}[h]
\centering
\includegraphics[page = 11, width=0.95\columnwidth]{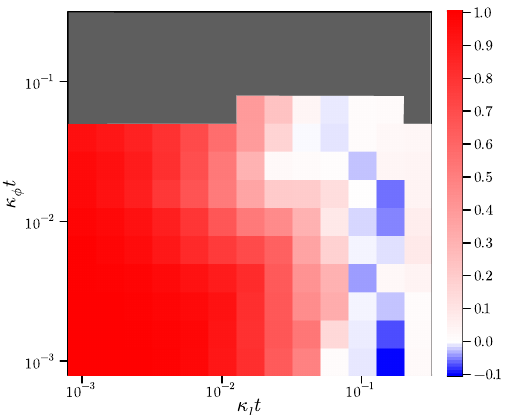} 
\caption{Percent difference in fidelity between the optimal \bino\ code and best \tworand\ code for each noise channel $(F_\bino - F_\tworand)/\max\{F_\bino, F_\tworand \}$. All the regions in blue represent noise channels where a random code outperformed binomial codes and all regions in grey denote regions where neither code beat the trivial code. This demonstrates that the random codes we implemented are most useful in a regime of high loss and moderate dephasing. Note: the scaling on the blue and red colors is unequal, this is necessary for visual clarity.
}\label{fig:compare_rand_bin}
\end{figure}

There is another interesting trend that is most obvious by comparing the rotation symmetry to the channel fidelity for cat and binomial codes.
Rotation codes have a well known tradeoff between code distance or ``protection'' against number errors which is $d_n = N$ and the code distance for phase errors $d_\phi = \pi/N $ such that 
\begin{align}\label{eq:num_phase_tradeoff}
d_n d_\phi &= \pi
\end{align} 
where $d_n-1$ is the number of loss or gain errors can be detected~\cite{GrimCombBara20,Yingkai2021,MarinoffBushCombes2023}.
This tradeoff is apparent in cat and binomial codes of \cref{fig:code_compare_and_sweep} where we observe that, for high dephasing, $N=2$ codes perform the best. As dephasing decreases, the rotation symmetry increases. It is important to note that we have not included $N=1$ codes or the \triv\ code. This trend seems to be much weaker in \onerand\ and \tworand\ codes, which again, could be due to insufficient sampling.

In \cref{fig:compare_rand_bin} we plot the difference in channel fidelity between the best binomial code and the best \tworand\ code. Specifically we plot 
\begin{equation}
\frac{ F_\bino - F_\tworand} {\max\{F_\bino, F_\tworand \} }
\end{equation}
The gray region indicates when the trivial code performs the best. When the plot is red, binomial codes beat \tworand\ codes, which is generally the case. When the plot is blue, \tworand\ codes beat binomial codes. For loss around $1\times 10^{-1}$, we see several points where \tworand\ codes outperform binomial codes. The improvement is modest, we see at most a $10\%$ improvement over binomial codes. There is a trade off between the amount of time spent searching for exceptional random codes and code performance. Here we focus on exploring trends across many noise channels with a relatively small number of samples per noise channel. The relative performance of random codes would increase with more sampling, but it is unclear if this performance would increase uniformly across all channels.

In \cref{fig:compare_rand_bin_XXX} of \cref{apx:more_plots}, we compare \cat\ and \bino\ codes in the same way. Interestingly, when we compare \cat\ to \tworand\ codes we find a larger region where \tworand\ codes outperform \cat\ codes.

\begin{figure}[!ht]
    \centering
    \includegraphics[page = 9, width=0.99\columnwidth]{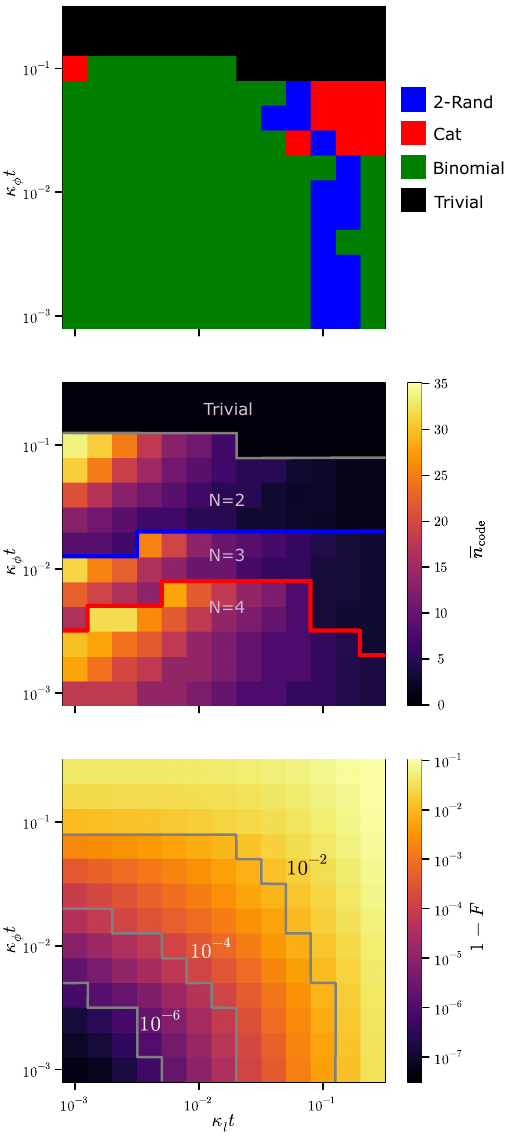}
    \caption{
    Best rotation code as a function of loss $\kappa_l t$ and dephasing  $\kappa_\phi t$ noise. 
    {\bf Row 1.} The best performing rotation code as a function of loss $\kappa_l t$ and dephasing  $\kappa_\phi t$. We note that \onerand\ codes were never optimal in our investigation.
    {\bf Row 2.} Average code space photon number of the best performing code as a function of loss and dephasing. The best rotation code symmetry values are also delineated.
    {\bf Row 3.} Channel infidelity of the best performing code.
    }
    \label{fig:best_of_all_codes}
\end{figure}

In \cref{fig:best_of_all_codes}, the best performing code in the set $\mathcal C \in [\triv, \cat, \bino, \onerand, \tworand]$ is found and plotted as a function of loss and dephasing. The first row 1 of the figure indicates which code is the best performing in the set of codes we looked at as a function of loss and dephasing. It is clear that \bino\ codes are typically the best performing rotation code that we investigated. That suggests most of the features observed on the subsequent plots will be similar to the binomial plots except in the large loss or large dephasing regime. This is evident in row 2 of \cref{fig:best_of_all_codes} shows similar striation to those in \cref{fig:code_compare_and_sweep} in photon number which are explain by changes in rotation symmetry of the optimal code. While the channel infidelity in row 3 will see modest improvements at large loss due to \tworand\ and \cat\ codes.

 
\section{Conclusions}\label{sec:conclusion}

We have introduced random rotation codes and systematically studied their performance as a function of loss and dephasing strength. By comparing the performance of these new codes to cat and binomial codes, we were able to find a region in our parameter space where random codes outperformed cat and binomial codes. Such results were found with relatively few samples, but it is expected that, with enough samples, random rotation codes could be at least as good as any known rotation code. Thus the numerics presented here are a proof of concept, and we hope it encourages further research using random bosonic codes. 

As a byproduct of our study on random codes, we have produced the first systematic numerical study of cat and binomial codes, which has revealed some interesting trends. The rotation symmetry, $N$, is strongly correlated with dephasing. This can be explained by the well known trade off between protection for phase errors and protection for number shift errors~\cite{Michael16,Yingkai2021,MarinoffBushCombes2023}. In particular, as dephasing decreases, the rotation symmetry of the best code increases. Moreover, the binomial code was the code that took up most of the ``phase diagram'' in \cref{fig:best_of_all_codes} of the optimal code for a given noise strength. This is to be expected as binomial codes were specifically designed for loss and dephasing errors.

We are presently focusing on exploring various notions of random translation codes and random bosonic codes in general. However, we think there are several interesting open questions that remain. Perhaps the most tantalizing would be changing the distribution of states that is being sampled. There is no reason to believe choosing codewords from the Haar measure will result in good codes. In fact this article provides evidence that they are not. A better idea might be to sample distributions that close to known good codes.\\

\noindent {\em Acknowledgments:} The authors acknowledge helpful discussions with Joshua Grochow, Rebecca Morrison, and Orit Peleg. AK, NL, and JC were supported by the Army Research Office through W911NF-23-1-0376 and National Science Foundation through a CAREER award ECCS-2240129 and Quantum Leap Challenge Institutes (QLCI) award OMA-2016244.

\bibliography{refs.bib}
\vspace{2em}

\newpage\newpage
\appendix

\begin{widetext}

\section{Additional plots for code comparison} \label{apx:more_plots}

\begin{figure*}[!ht]
    \centering
    \includegraphics[page = 7,width=0.99\textwidth]{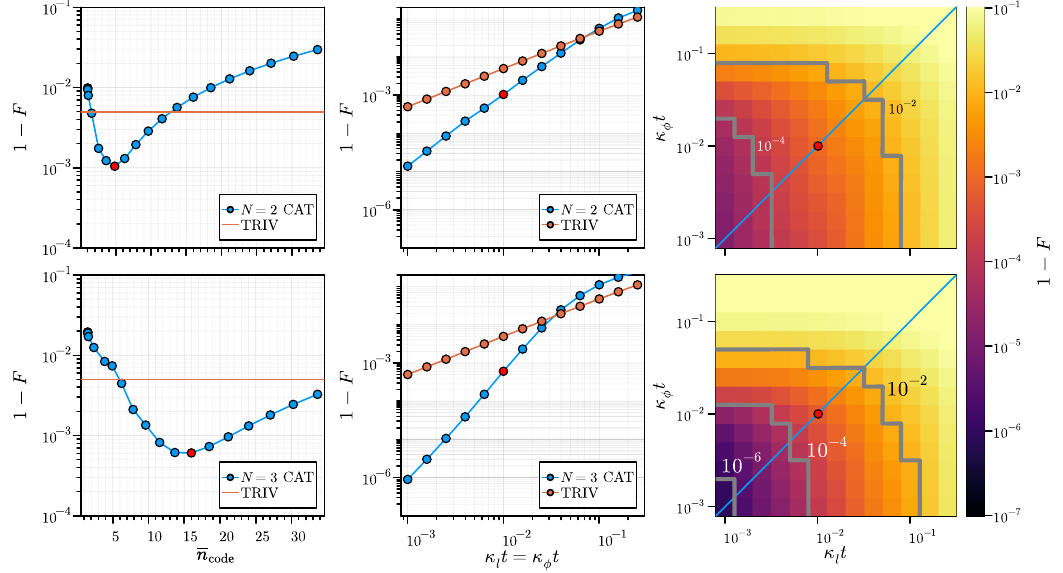}
    \caption{
    Performance of $N=2$ ({Row 1}) and $N=3$ ({Row 2}) cat codes.
    {\bf Column 1.} The performance (logical channel infidelity) of cat codes as a function of the average photon number of the code space projector when  $ \kappa_l t = \kappa_\phi t = 10^{-2}$. The minimum of this curve is plotted as a single red marker. 
    {\bf Column 2.} Code performance as a function of equal loss  and dephasing $\kappa_l t = \kappa_\phi t$. Notice that the minimum infidelity from column 1 is plotted as a red marker for $ \kappa_l t = \kappa_\phi t = 10^{-2}$.
    {\bf Column 3.} Code performance as a function of dephasing and loss. Note that the diagonal part of this plot is the data from column two where $\kappa_l t = \kappa_\phi t$.   
    }    \label{fig:cat_N2_N3_code_compare}
\end{figure*}

\begin{figure}[h]
\centering
\includegraphics[page = 10,width=0.99\columnwidth]{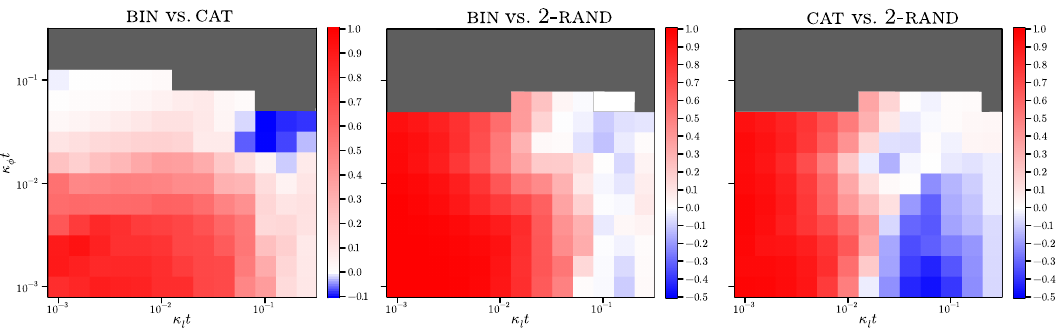} 
\caption{Difference in infidelity between codes as a function of loss and dephasing $(F_{\rm code 1} - F_{\rm code 2})/\max\{F_{\rm code 1}, F_{\rm code 2} \}$. The regions in blue represent noise channels where ``code 2'' outperformed ``code 1''. Grey regions are when no code beats the trivial encoding. Be mindful that the color bar scale changes from plot to plot.
{\bf Column 1.} $F_\bino - F_\cat$.
{\bf Column 2.} $F_\bino - F_\tworand$.
{\bf Column 3.} $F_\cat - F_\tworand$. Notice here that random codes can have $50\%$ improvement over cat codes in some cases. 
}\label{fig:compare_rand_bin_XXX}
\end{figure}

\end{widetext}

\end{document}